\documentclass[aps,pra,showpacs,twocolumn,superscriptaddress]{revtex4}

\usepackage{graphicx,color}
\usepackage{amsmath,amsthm,amsfonts,amssymb,bm}
\usepackage{times}
\usepackage{epsf}
\usepackage[colorlinks={true}]{hyperref}
\usepackage{bbold}
\usepackage{ulem}

\usepackage[T1]{fontenc}
\usepackage[utf8]{inputenc}
\usepackage{subfigure}
\usepackage{ulem}
\usepackage{verbatim}

\hypersetup{citecolor={blue}, filecolor={blue}, linkcolor={blue}, urlcolor={blue}}

\newcommand{\commentold}[1]{}
\DeclareMathSymbol{:}{\mathpunct}{operators}{"3A}

\begin{document}

\title{Inequivalence of Correlation-based Measures of Non-Markovianity}

\author{Alaor Cervati Neto}
\affiliation{Faculdade de Ci\^encias, UNESP - Universidade Estadual Paulista, Bauru, SP, 17033-360, Brazil}
\author{G\"{o}ktu\u{g} Karpat}
\affiliation{Faculdade de Ci\^encias, UNESP - Universidade Estadual Paulista, Bauru, SP, 17033-360, Brazil}
\author{Felipe Fernandes Fanchini}
\email{fanchini@fc.unesp.br}
\affiliation{Faculdade de Ci\^encias, UNESP - Universidade Estadual Paulista, Bauru, SP, 17033-360, Brazil}

\date{\today}

\begin{abstract}

We conclusively show that the entanglement- and the mutual information-based measures of quantum non-Markovianity are inequivalent. To this aim, we first analytically solve the optimization problem in the definition of the entanglement-based measure for a two-level system. We demonstrate that the optimal initial bipartite state of the  open system and the ancillary is always given by one of the Bell states for any one qubit dynamics. On top of this result, we present an explicit example dynamics where memory effects emerge according to the mutual information-based measure, even though the time evolution remains memoryless with respect to the entanglement-based measure. Finally, we explain this disagreement between the two measures in terms of the information dynamics of the open system, exploring the accessible and inaccessible parts of information.

\end{abstract}


\maketitle

\section{Introduction}

The study of open quantum systems has attracted much attention in the recent literature, which is mainly due to the increasing interest in the practical applications of quantum theory such as quantum information and communication protocols, quantum algorithms and quantum cryptology \cite{nielsenbook}. For potential quantum devices to effectively perform quantum information tasks in realistic conditions, they should be regarded and investigated as open quantum systems, i.e., their interactions with the surrounding environment should be carefully taken into account. The theory of open quantum systems in fact provide the necessary framework to examine the consequences of this interaction between a principal open system of interest and its surrounding environment \cite{breuerbook}. In general, such an interaction cannot be fully controlled and thus not desired since it typically destroys the valuable quantum properties in the principal system, due to the effects of decoherence.

From the viewpoint of memory effects, the time evolution of open quantum systems can be divided in two categories: Markovian and non-Markovian. While the absence of dependence of the quantum system on its past time evolution implies a Markovian and thus a memoryless quantum process, non-trivial temporal correlations among different states of the system throughout the dynamics give rise to a non-Markovian quantum process exhibiting memory effects. Indeed, the presence of memory in the dynamics can help protecting coherence and precious quantum correlations in open systems for prolonged time intervals \cite{entnonmark}. Therefore, understanding the nature of memory effects from various perspectives has become a significant problem for the open quantum systems community. Numerous different techinuques have been put forward to measure the degree of memory effects \cite{nmreview1,nmreview2,rivas10, hou11, lu10, luo12, bylicka14, chr14, breuer09, bogna16}. However, it is known that non-Markovianity in quantum physics is a many-sided phenomenon and different methods for quantifying memory effects do not agree with each other in general \cite{notagree1,notagree2}, as they recognize different aspects of the memory.

The distribution and flow of information inside the open system and its reservoir throughout the dynamics plays a very important role in the creation of non-Markovian memory effects. A natural intuition is that if the information flows from the open system to the environment in a monotonic fashion in course of the time evolution, the resulting dynamics can be regarded as Markovian due to the lack of memory effects. However, if there exits a temporary back-flow of information from the environment to the open system, then memory effects emerge since future states of the open system can now depend on its past states. In fact, one of the first quantifiers of non-Markovianity has been proposed based on the idea of information back-flow \cite{breuer09}. In that approach, the variations in the distinguishability of an arbitrary pair of initial system states, measured with the help of trace distance, are interpreted as information flow between the system and the environment. 

On the other hand, there are two other well-known characterizations of non-Markovian memory effects which are defined via the dynamics of correlations (namely, entanglement and mutual information) in bipartite states that are composed of a principal open system and an additional ancillary \cite{rivas10,luo12}. Both of these two distinct measures have also been recently revealed to be very closely related to the flow of information, quantified via entropic quantities, between the system and its environment \cite{fanchini14,haseli14,karpat15}. All the same, analytical evaluation of almost all measures of non-Markovianity is difficult due to the optimization procedures required to remove the dependence of the measures on initial conditions, for non-Markovianity is a property of quantum processes and not specific states.

In this work, we consider the two correlation-based measures of quantum non-Markovianity, namely entanglement- and mutual information-based measures, both of which can be individually related to the information theoretic understanding non-Markovian memory effects in terms of the back-flow of information. We commence by presenting a simple analytical argument that solves the optimization problem in the definition of the entanglement-based measure for a two-level system. In particular, we prove that, in the space of pure states, the optimal initial state of the open system and ancilla is always one of the Bell states, independently of the specific open system dynamics. Then, based on this result, we conclusively demonstrate through an explicit example that the entanglement- and mutual information-based measures are fundamentally inequivalent. Lastly, we provide an information theoretic explanation for this disagreement in terms of the concepts of accessible and inaccessible information.

This paper is organized as follows. In Sec. II, we introduce the quantifiers of memory effects that we intend to explore in this work, and also the solution of the optimization problem for the entanglement-based measure. Sec. III presents an explicit example demonstrating the inequivalence of the two correlation-based measures of non-Markovianity. In Sec. IV, we give an information theoretic explanation accounting for this inequivalence. Sec. V includes our conclusion.

\section{The degree of memory effects}

In this section, we intend to briefly review the definitions of the entanglement- and mutual information-based measures of non-Markovianity, and discuss the optimization problem for the former measure in case of a two-level system dynamics.

Let us first describe the basics of how one can identify the non-Markovian memory effects and quantify their degree using the correlations. Suppose that we have a quantum process $\Lambda$, i.e., a completely positive trace preserving (CPTP) map, that represents the dynamical evolution of a quantum system $S$ in the state $\rho_S$. In addition, we introduce an trivially evolving ancillary system $A$ in the state $\rho_A$ having the same dimension as $S$, so that the composite system of the open system and the ancillary is in the state $\rho_{SA}$. We note that the memoryless Markovian maps satisfy the decomposition relation $\Lambda(t,0)=\Lambda(t,s)\Lambda(s,0)$, where $\Lambda(t,s)$ is a CPTP map with $s \leq t$. In other words, the dynamical map $\Lambda$ is a divisible map implying a memoryless time evolution for the open quantum system. Recalling the fact that either the amount of entanglement $E$ or the quantum mutual information $I$ can increase under local CPTP maps, and the dynamical map $\Lambda$ acts only on the subsystem $S$, memoryless evolution implies that
\begin{align}
E(( \Lambda(t,0)\otimes\mathbb{1})\rho_{SA}) \leq& E(( \Lambda(s,0) \otimes\mathbb{1})\rho_{SA}), \\[4pt]
I(( \Lambda(t,0)\otimes\mathbb{1})\rho_{SA}) \leq& I(( \Lambda(s,0) \otimes\mathbb{1})\rho_{SA}),
\end{align}
at all times $0\leq s \leq t$ for all composite quantum states $\rho_{SA}$, and $\mathbb{1}$ is the identity operation. Hence, it is possible to identify the presence of non-Markovian memory effects through the violation of one of the above inequalities. Specifically, any revival of entanglement $E$ or quantum mutual information $I$ during the dynamics can be regarded as a signal of the existence of memory effects. Moreover, based on the amount of the violation of these inequalities, we can define two distinct measures for the degree of memory as \cite{rivas10,luo12}
\begin{align}
\mathcal{N}_{E}(\Lambda)=&\max_{\rho_{SA}}\int_{(d/dt)E_{SA}>0}\frac{d}{dt} \label{entbas}
E_{SA} dt, \\[3pt]
\mathcal{N}_{I}(\Lambda)=&\max_{\rho_{SA}}\int_{(d/dt){I}_{SA}>0}\frac{d}{dt}{I_{SA}}dt, \label{mutbas}
\end{align}
where $E_{SA}=E((\Lambda\otimes\mathbb{1})\rho_{SA})$ and $I_{SA}=I((\Lambda\otimes\mathbb{1})\rho_{SA})$ and the maximization is over all possible pure initial states of the bipartite system $\rho_{SA}$. We emphasize that, although quite similarly defined, entanglement- and mutual information-based measures are distinct since they are proposed as individual witnesses for non-divisibility, and they can signal the presence of memory independently of each other. Additionally, it is critical to note that even though they have been originally introduced without any operational interpretation in terms of information flow between the system and its environment, they have been very recently shown to be used to establish information theoretic definitions of non-Markovianity \cite{fanchini14,haseli14,karpat15}.

\subsection*{Optimization of the entanglement-based measure}

Since the evaluation of both measures requires the solution of a non-trivial optimization problem, it is no simple task to prove that they are actually inequivalent. In order to conclusively show their inequivalence, it is necessary to solve the optimization problem for at least one of them. Here, we will present an analytical solution to this problem for the entanglement-based measure of quantum non-Markovianity given in Eq. (\ref{entbas}), i.e., we will demonstrate that the global maximum for the entanglement-based measure is always reached by the Bell state $|\Phi^+\rangle$ for any single qubit dynamics.

Let us first recall the well-known entanglement evolution equation \cite{konrad}, which is constructed upon the celebrated Choi-Jamiolkowski isomorphism between the states and the quantum maps. Konrad et al. have proven that, given a pure bipartite initial state $\rho_{SA}$ and an arbitrary dynamical map $\Lambda$, the evolution of entanglement as measured by concurrence \cite{conc} can be described by the following equation
\begin{equation} \label{entev}
C[(\Lambda \otimes\mathbb{1})\rho_{SA}]=C[(\Lambda\otimes \mathbb{1})|\Phi^+\rangle\langle\Phi^+|]C(\rho_{SA})
\end{equation}
where $|\Phi^+\rangle=(|00\rangle+|11\rangle)/\sqrt{2}$, and $C(\rho_{SA})$ is the concurrence of the initial state. $C[(\Lambda\otimes\mathbb{1})\rho_{SA}]$ and $C[(\Lambda\otimes\mathbb{1})|\Phi^+\rangle\langle\Phi^+|]$ are respectively the concurrence of the pure initial state $\rho_{SA}$ and the Bell state $|\Phi^+\rangle\langle\Phi^+|$ as the environment interacts with the open system $S$. Thus, the concurrence of all pure states evolve as the maximally entangled state, except for the fact that their entanglement is rescaled by the amount of entanglement in the initial state. Next, if we take the time derivative of the both sides of the above equation we have
\begin{equation}
\frac{d}{dt}C[(\Lambda\otimes\mathbb{1})\rho_{SA}]= C(\rho_{SA}) \times \frac{d}{dt}C[(\Lambda\otimes\mathbb{1})|\Phi^+\rangle\langle\Phi^+|].
\end{equation}
As can be clearly seen, a similar relation holds for the rate of change of the concurrence. That is to say that the rate of change of the evolution of concurrence for any pure initial state is exactly as that of a maximally entangled state, rescaled by the amount of entanglement in the initial state. Hence, this relation proves that the greatest amount of revival in the concurrence throughout dynamics of the open system is always attained for the maximally entangled initial state.

As we plan to utilize entanglement of formation in our later discussions to measure the entanglement for our purposes, we should mention that the optimality of the maximally entangled Bell state $|\Phi^+\rangle$, for the entanglement-based non-Markovianity measure in Eq. (\ref{entbas}), is not limited to the case where we use concurrence as the measure of entanglement. In fact, entanglement of formation and concurrence are closely related to each other, the former being a monotonically increasing function of the latter \cite{eof}. Consequently, each minimum and maximum in the dynamics of the entanglement necessarily occur simultaneously for both measures, which proves that the Bell state $|\Phi^+\rangle$ is also the optimal initial state for Eq. (\ref{entbas}) in case we use entanglement of formation to quantify entanglement.
 
\begin{figure}[t]
\includegraphics[width=0.33\textwidth]{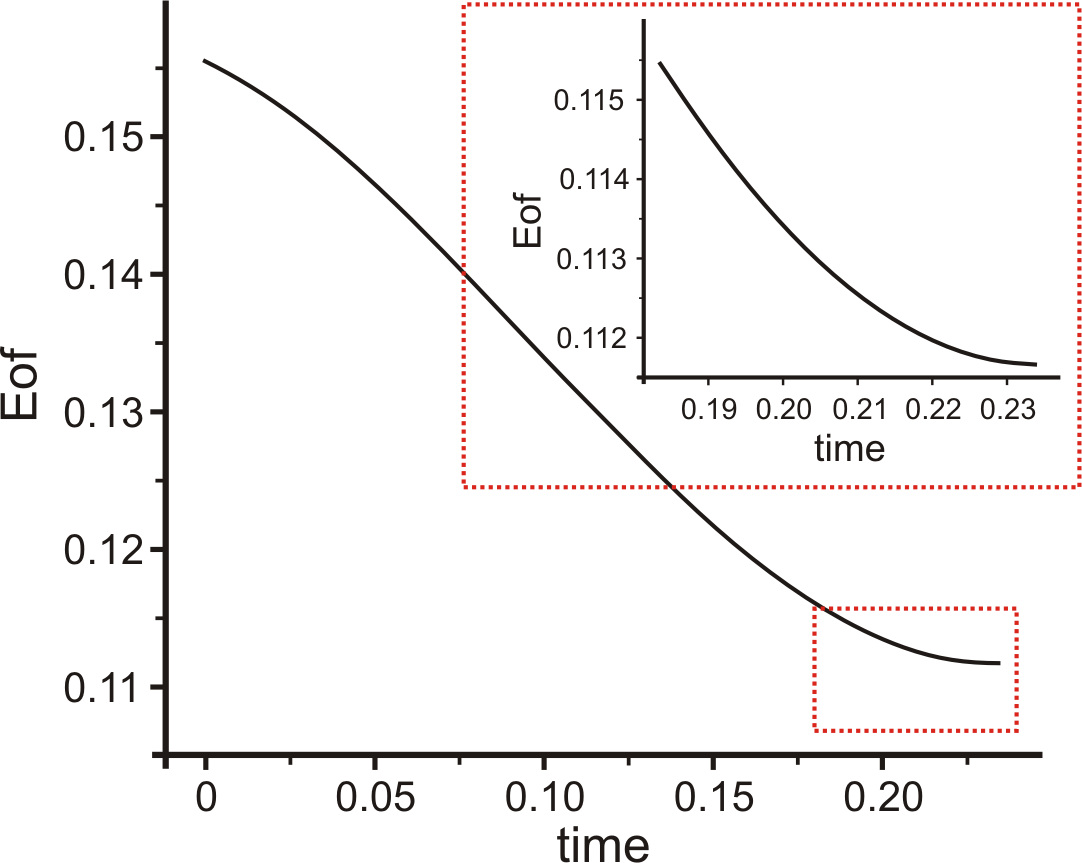}
\caption{(Color Online) The entanglement of formation (Eof) versus the dimensionless time for the considered amplitude damping model with $t_c=0.25$. The inset figure shows a zoomed in version of the region of our interest.}\label{fig1}
\end{figure} 
 
\vspace{-0.1cm}
 
\section{Inequivalence of the non-Markovianity Measures: An Example Dynamics}

Now that we have introduced the correlation-based quantifiers of non-Markovian memory effects, we are in a position to present an open quantum system model using which we can clearly demonstrate the inequivalence of the non-Markovianity measures. We should first note that the solution of the optimization problem for the mutual information-based measure in Eq. (\ref{mutbas}) is not known in general, and in fact, the optimal initial state has been shown to be model dependent \cite{notagree1}. However, since we have solved the optimization problem for the entanglement-based measure, it is sufficient for us to find a open system model for which the quantum process is Markovian according to the entanglement-based measure while it is non-Markovian according to the mutual information-based measure. More specifically, the generalized amplitude damping model we will consider here induces temporary revivals in the mutual information but not in the entanglement throughout the time evolution of the open quantum system. 

In order to be able to evaluate the non-Markovianity measures $\mathcal{N}_{E}$ and $\mathcal{N}_{I}$, we need to calculate the time evolution of the composite system of the open system and the ancillary, where the latter evolves trivially in time. In particular, the dynamics of the composite system $\rho_{SA}$ under the generalized amplitude damping model can be expressed with the help of the operator-sum representation in the following way:
\begin{equation}
\rho_{SA}(t)=\sum_iK_{i}(t)\rho_{SA}(0)K_{i}^\dagger(t)\label{dynamics}
\end{equation}
where the Kraus operators describing the dynamics read
\begin{align}
	K_{1}\left(t\right)&=\sqrt{s\left(t'\right)}\left(\begin{array}{cc}1&0\\0&\sqrt{r\left(t'\right)}\end{array}\right)\otimes\mathbb{1}, \\
	K_{2}\left(t\right)&=\sqrt{s\left(t'\right)}\left(\begin{array}{cc}0&\sqrt{1-r\left(t'\right)}\\0&0\cr\end{array}\right)\otimes\mathbb{1},\nonumber\\
	K_{3}\left(t\right)&=\sqrt{1-s\left(t'\right)}\left(\begin{array}{cc}\sqrt{r\left(t'\right)}&0\\0&1\cr\end{array}\right)\otimes\mathbb{1},\nonumber\\
	K_{4}\left(t\right)&=\sqrt{1-s\left(t'\right)}\left(\begin{array}{cc}0&0\\ \sqrt{1-r\left(t'\right)}&0\cr\end{array}\right)\otimes\mathbb{1},\nonumber
\end{align}
with $s\left(t'\right)=\cos^{2}\omega{t'}$ and $r\left(t'\right)=e^{-t'}$. Here we have $t'\equiv t-H_{s}\left(t-t_{c}\right)\left(t-t_{c}\right)$, where $H_s(x)$ is the Heaviside step function defined as $H_s(x)=\frac{d}{dx}\max\{x,0\}$, and $t_c$ is the critical time after which the decoherence is ceased and the dynamics becomes trivial. In other words, we consider the case where the open quantum becomes protected from the effects of the environment once a critical instant $t_c$ is reached. The reason we construct the model in this way is for the sake of simplicity, since we can already prove our point without the need for considering the full decoherence dynamics.

\begin{figure}[t]
\includegraphics[width=0.335\textwidth]{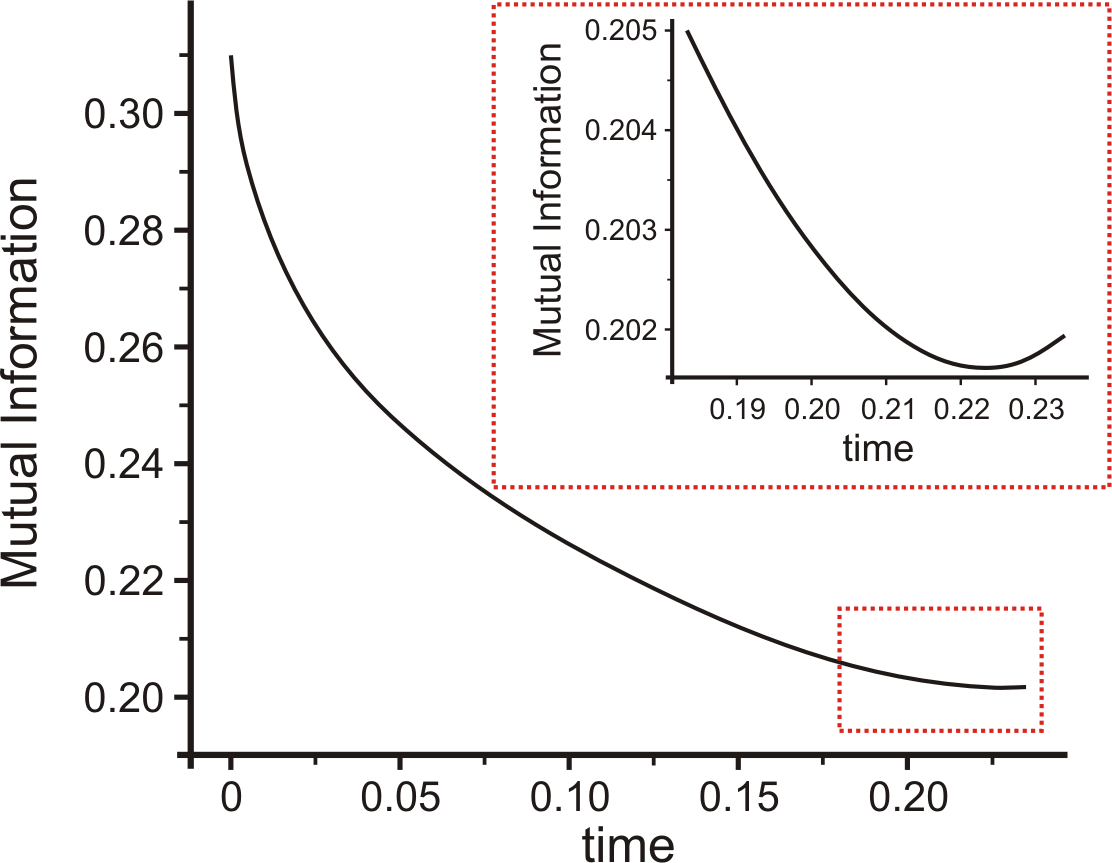}
\caption{(Color Online) The mutual information versus the dimensionless time for the considered amplitude damping model with $t_c=0.25$. The inset figure shows a zoomed in version of the region of our interest.}\label{fig2}
\end{figure}

\begin{figure*}[t]
\includegraphics[width=0.95\textwidth]{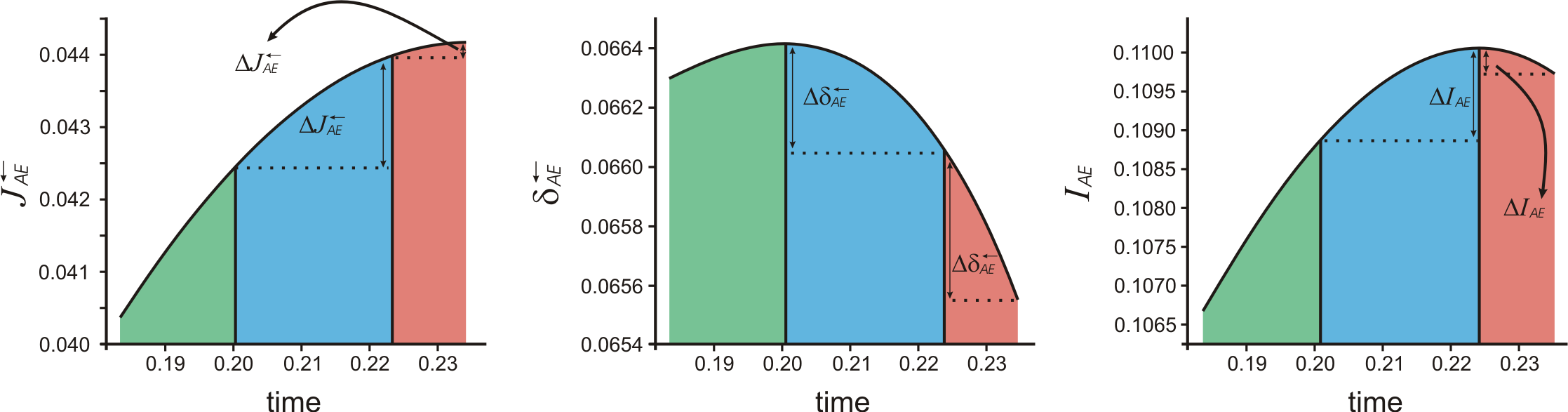} 
\caption{(Color Online) The accessible information $J_{AE}^\leftarrow$, the inaccessible information $\delta_{AE}^\leftarrow$, and the quantum mutual information $I_{AE}$ (sum of the accessible and inaccessible parts) versus the dimensionless time for the time interval of our interest for the amplitude damping model.}\label{fig3}
\end{figure*}

In Fig. \ref{fig1} and Fig. \ref{fig2}, we respectively display the time evolution of the entanglement-based measure $\mathcal{N}_{E}$ and the mutual information-based measure $\mathcal{N}_{I}$ for the generalized amplitude damping channel with the critical time $t_c=0.25$ and $\omega=5$. While the initial state of the system and the ancillary $\rho_{SA}$ is chosen as the Bell state $|\Phi^+\rangle$ for the entanglement-based measure, in case of the mutual information-based measure, we chose $\rho_{SA}=|\Psi\rangle\langle\Psi|$ where $|\Psi\rangle = a|\!\uparrow\uparrow\rangle + b|\!\uparrow\downarrow\rangle + c|\!\downarrow\uparrow\rangle + d|\!\downarrow\downarrow\rangle$ with $a=0.05$, $b=0.95$, $c=0.17$, and $d=\sqrt{1-a^2 - b^2 - c^2}$. As we proved that $|\Phi^+\rangle$ is the optimal state for the entanglement-based measure, we conclude that the quantum process is definitely Markovian with respect to it since there is no revival in the dynamics of entanglement. Nonetheless, we observe that the dynamics of mutual information exhibits a revival where entanglement remains monotonically decreasing, which can be clearly seen in the inset of Fig. \ref{fig2}. Thus, the considered example conclusively shows that the measures $\mathcal{N}_{E}$ and $\mathcal{N}_{I}$ are inequivalent in general.

\section{Explaining the inequivalence: flow of accessible and inaccessible information}

In this section, we aim to investigate the roots the inequivalence of the two considered correlation-based measures of quantum non-Markovianity using an information theoretic approach. In order to be able to understand the difference between the two measures, we first need to review a few basic concepts from the quantum information theory.  
 
The accessible information, also known as the classical correlation, quantifies the maximum amount of classical information that can be extracted about one subsystem by performing local observations on another one \cite{hend}. In particular,
\begin{equation}
J_{AE}^\leftarrow= \max_{\{\Gamma_{i}^{E}\}} \left[S(\rho_{A}) - \sum_i p_i (\rho_{A}^i|\Gamma_i^{E})\right],\label{ai}
\end{equation}
measures the maximum amount of classical information that the environment $E$ can obtain about the ancillary system $A$. Here $\{\Gamma_i^{E}\}$ is a complete positive operator valued measure (POVM) acting on the environment $E$, and $\rho_{A}^{i}=\textmd{Tr}_{E}((\mathbb{1}^{A}\otimes\Gamma_{i}^{E})\rho_{AE})/p_{i}$ is the  state of ancillary subsystem $A$ after obtaining the outcome $i$ with the probability $p_{i}=\textmd{Tr}((\mathbb{1}^{A}\otimes\Gamma_{i}^{E})\rho_{AE})$. On the other hand, the inaccessible information quantifies the minimum amount of quantum information that cannot be extracted about one system by locally observing the other. As quantum mutual information is defined as the total amount of quantum and classical correlations that two subsystems share, then the inaccessible information reads
\begin{equation}
\delta_{AE}^\leftarrow = I_{AE} - J_{AE}^\leftarrow,
\end{equation}
which is in fact nothing but the famous quantum discord \cite{olli}.

Let us additionally remember how the entanglement- and the mutual information-based measures of non-Markovianity are respectively related to the flow of accessible and total information in the composite system of open system, ancilla and the environment. Provided we assume that the tripartite state of the whole system $SAE$ is initially pure, it can be shown that \cite{fanchini14} the monotonically decreasing behaviour of the entanglement of formation in the bipartite system $SA$ is directly liked with the monotonically increasing behaviour of the accessible information in the bipartite system $AE$ as 
\begin{equation}
\frac{dE_{SA}}{dt} = -\frac{dJ_{AE}^\leftarrow}{dt}.\label{dEdt}
\end{equation}
To put it differently, if the amount of accessible information that the environment $E$ can obtain about the ancilla $A$, by means of an interaction with $S$, temporarily decreases during the dynamics, then the memory effects emerge and the process becomes non-Markovian based on Eq. (\ref{entbas}). In fact, a similar relation holds between the rate of changes of the mutual information in the bipartite systems $SA$ and $AE$, 
\begin{equation}
\frac{dI_{SA}}{dt} = -\frac{dI_{AE}}{dt} = -\left(\frac{dJ_{AE}^\leftarrow}{dt}+\frac{d\delta_{AE}^\leftarrow}{dt}\right),\label{dIdt}
\end{equation}
which links the emergence of non-Markovian memory effects, with respect to Eq. (\ref{mutbas}), to the temporary decrease of the total correlations (accessible and inaccessible) in the bipartite state $AE$ throughout the dynamics of the open system \cite{haseli14}. Hence, comparing Eq. (\ref{dEdt}) to Eq. (\ref{dIdt}), it becomes rather straightforward to see that the equivalence or inequivalence of the two correlation-basd measures of non-Markovianity is determined by the rate of change of the inaccessible information $\delta_{AE}^\leftarrow$.

Before starting to investigate the reason behind the inequivalence of the measures, we should stress that the above discussion, regarding the distribution of correlations in the total system $SAE$, holds under the assumption of a zero-temperature environment, i.e., in case we have a pure initial state for the environment $E$. All the same, even if we have an open system model where the environment starts in a mixed state (finite temperature environment), it can be purified without loss of generality and thus the same conclusions based on Eq. (\ref{dEdt}) and Eq. (\ref{dIdt}) can still be drawn by replacing $E$ by $EE'$, where $E'$ is an additional purifying environmental system.

In order to comprehend the relation between the two measures, we plot in Fig. \ref{fig3} the time evolution of the accessible information $J_{AE}^\leftarrow$, the inaccessible information $\delta_{AE}^\leftarrow$, and the quantum mutual information $I_{AE}$ (sum of the accessible and inaccessible information) for the considered amplitude damping model. We only show the dynamical behaviour of these three quantifiers for a specific time interval where the inequivalence of the correlation-based measures becomes evident. For a better explanation we divide the dynamics in all three plots in three regions, as can be clearly distinguished in Fig. \ref{fig3}. In the first green region, all of the considered quantities $J_{AE}^\leftarrow$, $\delta_{AE}^\leftarrow$, and $I_{AE}$ increase monotonically which implies a Markovian quantum process according to both measures. In the second blue region, $J_{AE}^\leftarrow$ keeps increasing monotonically pointing out to Markovian behavior with respect to the entanglement-based measure. We also note that although the rate of change of the inaccessible information $\delta_{AE}^\leftarrow$ becomes negative in this region, the mutual information $I_{AE}$ preserves its increasing trend since the magnitude of the rate of change of the accessible information is greater than that of the inaccessible information, that is, $|dJ_{AE}^\leftarrow/dt|>|d\delta_{AE}^\leftarrow/dt|$. Finally, in the third red region,  both the accessible information $J_{AE}^\leftarrow$ and the inaccessible information $\delta_{AE}^\leftarrow$ behaves in the same way as in the second region, i.e., $J_{AE}^\leftarrow$ keeps increasing and $\delta_{AE}^\leftarrow$ decreasing monotonically. However, there is one crucial difference here, which is the fact that the magnitude of the rate of change of the inaccessible information is now greater than that of the accessible information, that is, $|d\delta_{AE}^\leftarrow/dt|>|dJ_{AE}^\leftarrow/dt|$. As a consequence, the mutual information $I_{AE}$ starts decreasing implying a non-Markovian dynamics, whereas the dynamics is still completely Markovian based on the behaviour of entanglement of formation. Indeed, this outcome demonstrates that the inequivalence of the entanglement- and mutual information-based measures is strictly related with the balance between the decay and growth of accessible and inaccessible parts of the information. In other words, if the environment loses quantum information faster than it gains classical information, then the correlation-based measures of non-Markovianity are inequivalent.    

Before concluding our study, we would like to comment on the possible extension of the analysis we have presented here to the case of different types of noise, namely, the phase damping noise scenario. Indeed, we have also explored several open system models describing dephasing type dynamics such as the well-known non-Markovian coloured dephasing model first introduced by Daffer et al. in Ref. \cite{daffer}. Although we cannot claim that this is a general result, we have found for phase damping models that both the entanglement- and the mutual information-based measures of non-Markovianity leads to the same conclusion when they are used to determine whether the dynamics is Markovian or not. In other words, we have not been able to find an example dephasing dynamics where the optimized entanglement-based measure characterizes a Markovian evolution whereas the mutual information-based measure determines a non-Markovian evolution. Thus, in case of dephasing, it seems like the magnitude of the rate of change of the accessible information $|dJ_{AE}^\leftarrow/dt|$ is greater than the magnitude of the rate of change of the inaccessible information $|d\delta_{AE}^\leftarrow/dt|$, which makes both correlation-based measures equivalent. Therefore, it is still an open problem to prove or disprove that the two considered correlation-based non-Markovianity measures are always equivalent in case of dephasing noise. Lastly, one can also think about considering the generalization of our discussions to two-qubit dynamics. However, such a problem turns out to be extremely difficult  due to the fact that the analytical solution of the optimization problem for either measures is no longer available and also the structure of entanglement in higher dimensions is quite complex.

\section{Conclusion}

We have presented a conclusive proof that the correlation-based measures of quantum non-Markovianity are inequivalent in general. In case of single qubit quantum processes, we have demonstrated that the optimal initial bipartite state of the open system and the ancilla for the entanglement-based measure is given by one of the maximally entangled Bell states, independently of the model for the open system dynamics. Based on this finding, we have proved the inequivalence of the entanglement- and mutual information-based measures through an explicit example dynamics that describes an amplitude damping type of process for a single two-level system. In addition, we have provided an information theoretic explanation for when different conclusions can be drawn from the two considered measures. Our treatment has revealed that when the environment loses quantum information to the ancillary system faster than it gains classical information from it, the two correlation-based measures of quantum non-Markovianity become inequivalent.

\begin{acknowledgments}
GK gratefully acknowledges the financial support from FAPESP under the grant number 2012/18558-5, and FFF under the grant number 2015/05581-7. FFF also acknowledges the support from CNPq under the grant number 474592/2013-8 and by INCT-IQ under the process number 2008/57856-6. 
\end{acknowledgments}

\end{document}